\documentclass[aps,prb,showpacs,twocolumn,groupedaddress,amsmath,amssymb,floatfix]{revtex4}
\usepackage{graphicx}
\usepackage{epsfig}
\begin{document}

%\preprint{}

\title{Scaling of the thermal resistivity of $^4$He in restricted geometries}


\author{Chongshan Zhang, Kwangsik Nho, and D. P. Landau}
%\thanks{}
\affiliation{Center for Simulational Physics\\The University of Georgia\\Athens, GA 30602}

\date{\today}

\begin{abstract}
The thermal resistivity and its scaling function in quasi-2D $^4$He systems are studied by Monte Carlo and spin-dynamics simulation. 
We use the classical 3D XY model on $L\times L\times H$ lattices with $L\gg H$,
applying open boundary condition along the $H$ direction and periodic boundary conditions along the $L$ directions. 
A hybrid Monte Carlo algorithm is adopted to efficiently deal with the critical slowing down and to produce initial states for time integration. 
Fourth-order Suzuki-Trotter decomposition method of exponential operators is used to solve numerically the coupled equations of motion for each spin. 
The thermal conductivity is calculated by a dynamic current-current correlation function. 
Our results show the validity of the finite-size scaling theory, and the calculated scaling function agrees 
well with the available experimental results for slabs using the temperature scale 
and the thermal resistivity scale as free fitting parameters.
\end{abstract}

%\keywords{}

\maketitle

\section{\label{sect:intro}Introduction}
High-resolution measurements of various properties of $^4$He confined in restricted geometries near the bulk superfluid transition temperature $T_{\lambda}$ have been extensively carried out for over three decades \cite{MG97, MKG99, KMG00, LSN00, SWD90, SD94, KD92, NCL94, WD94, SM94, SM95, SM295, KA95, NM01}. 
The measurements approach so close to $T_{\lambda}$ that the correlation length becomes macroscopic in size. As a result, the whole fluid acts in a correlated way and the values of global properties are changed. This offers an opportunity for testing the finite-size scaling theory which describes the effect of confinement in a finite geometry near a critical point.

Physical systems which exhibit a second-order phase transition and which are confined in a finite geometry (e.g., a film, a pore, or a box) are thought to be well described by the finite-size scaling theory\cite{MEF74} at temperatures close to the critical temperature $T_{\lambda}$. The finite-size scaling theory is based on the idea that the finite-size effects can be observed when the correlation length $\xi$ becomes of the order of the finite system size (i.e., the side of the cube, the thickness of the film, or the diameter of the pore). For a physical quantity $O$, this statement can be expressed as follows\cite{EB82}:
\begin{eqnarray}
\frac{O(t,H)}{O(t,H=\infty)}=f(x), \label{eq1}
\end{eqnarray}
where
\[
t=\frac{\mid T-T_{\lambda}\mid}{T_{\lambda}}
\mbox{\ \ \ and\ \ \ }
x=\frac{H}{\xi(t,H=\infty)}.
\]
%\begin{eqnarray}
%t=\mid T-T_{\lambda}\mid /T_{\lambda},\nonumber \\
%x=\frac{H}{\xi(t,H=\infty)}. \nonumber
%\end{eqnarray}
%\mbox{where \ \ \ \ } x=\frac{H}{\xi(t,H=\infty)}.
Here $H$ denotes the relevant confining length and $\xi(t, H=\infty)$ is the correlation length of the bulk system. $f(x)$ is a universal function which does not depend on the microscopic details of the system. It does, however, depend on the observable $O$, the geometry of the system, and boundary conditions imposed on the system.

Liquid helium $^4$He has been widely used to examine the validity of the finite-size scaling theory of critical phenomena; sophisticated experimental studies were carried out in microgravity environment, for example, Lipa $et$ $al.$\cite{JAL00} measured the specific heat of helium confined in a parallel plate geometry with a spectacular nanokelvin resolution, and the shape of the confining geometry, such as film, pore, and box, can be designed with such a precision that the relevant confining length is well defined.

For static properties, the scaling of superfluid $^4$He has been studied experimentally, analytically and numerically. For example, the specific heat near the superfluid transition of $^4$He has been measured for confinements which vary by a factor of over 1000 and the data of a large extent can be collapsed upon a unique function when properly reduced\cite{MG97,MKG99,KMG00,LSN00}. Field-theoretical calculations for the standard Landau-Ginzbrug free energy functional in different geometries with dirichlet boundary conditions have been carried out\cite{SWD90,SD94,KD92} and the results  agree with the specific heat measurements\cite{NCL94,WD94}. Schultka and Manousakis calculated the superfluid density and the specific heat of $^4$He in film and pore geometries using Monte-Carlo simulation and they demonstrated the validity of the finite-size scaling theory\cite{SM94,SM95,SM295}.

Beside static properties, finite-size scaling theory can be also understood by studying dynamical and transport properties near the critical point. Among them, the thermal conductivity $\lambda$ of $^4$He is a good candidate, because it is a measurable property and its bulk transition has been carefully studied\cite{TA85,DZM86,TA86, HS74}.
D. Murphy et al.\cite{DE03} and E. Genio et al.\cite{ED05} measured the thermal conductivity of liquid $^4$He confined in a "microchannel plates" of thickness 2 mm with holes 0.5 $\mu$m and 1 $\mu$m in diameter, and their data are consistent with a universal scaling function. Nho and Manousakis\cite{NM01} studied the thermal conductivity $\lambda$ of $^4$He confined in a pore-like geometry and their simulational results agree well with the experimental results.
Rather recently, Jerebets et al. measured the thermal conductivity of liquid $^4$He confined in a film geometry\cite{AL05}. In this paper we will study the thermal conductivity $\lambda$ of confined helium using Monte Carlo and spin-dynamics methods, calculate the scaling function associated with $\lambda$ for a film geometry, and compare it with the experimental results.

The remainder of the paper is organized as follows: In Sec. II, we will present the model, the simulation methods, and the method for extracting the thermal conductivity. In Sec. III, we will examine the finite-size scaling theory for the thermal resistivity and compare our calculated scaling function with experimental results. We summarize our results in the last section.

\section{\label{sect:model}MODEL AND SIMULATION METHOD}
In this section, we will briefly describe the model and simulation methods used to study the properties of superfluid $^4$He and show how the thermal conductivity is computed in our model. Matsubara and Matsuda\cite{MM56} have proposed a lattice model to explain the properties of liquid helium. In the model, the liquid is regarded as a lattice composed of atoms and holes. In terms of operators which create or annihilate an atom at each lattice point, it is proved that the lattice model is equivalent to the ½ spin system. Both systems are classified as model F (or E in the absence of an external field) in the classification of dynamical models and same universality class with the XY model, so we can use the XY model to study the properties of liquid $^4$He\cite{HH77,NM99}.

The Hamiltonian of the XY model on a lattice takes the following form:
\begin{eqnarray}
{\cal H}=-J\sum_{<ij>}(S_i^x S_j^x + S_i^y S_j^y),  \label{eq2}
\end{eqnarray}
where $J$ sets the energy scale,$<ij>$ denotes a nearest neighbor pair of spins on a simple lattice in three dimensions, and the summation is over all nearest neighbors. In this model, each spin is a classical spin with three components, $\vec{S_i}=(S_i^x, S_i^y, S_i^z)$.

In our calculations, we use a film geometry, i.e. a $L\times L\times H$ lattice with $L\gg H$, in order to mimic experiment. Open boundary conditions are imposed along the confining dimension (the $H$ direction). In our model the spins at the open boundary have no neighbors outside the confining space. Periodic boundary conditions are used along the large planar dimensions ( $L$ direction), because they approximate the limit $L\rightarrow \infty$ better.

The thermal conductivity $\lambda$ of liquid $^4$He at a given temperature $T$ can be calculated using the dynamic current-current correlation function\cite{KL99, LH74}
\begin{eqnarray}
\lambda \propto \int_{0}^{\infty} dt\sum_{i}<j_0^z(0)j_i^z(t)>,
\end{eqnarray}
where the $z$ component $j_i^z$ of the current density $\vec{j_i}$ associated with the lattice point $i$ is defined by
\begin{eqnarray}
j_i^z = J(S_i^yS_{i+e_z}^x - S_i^xS_{i+e_z}^y),     \label{eq4}
\end{eqnarray}
where the notation $i+e_z$ denotes the nearest neighbor of the lattice site $i$ in the $z$ lattice direction.

The thermal conductivity calculated by above equations strongly depends on the initial configuration of the system. The thermal conductivity at a given temperature $T$ is an average value for all possible initial configurations. Therefore, a sequence of uncorrelated equilibrium configurations is needed to provide starting points for the spin dynamics.

These configurations are obtained from a Monte Carlo simulation using the Hamiltonian given by Eq.(2). We use a hybrid Monte Carlo procedure\cite{KL99}, which consists of 10 updates (MCCMOCMCCO), where M is Metropolis update\cite{NAMAE53}, C is cluster update\cite{UW89} and O is  the over-relaxation algorithm\cite{BW87}.

Using this hybrid algorithm, we generate approximately 2000 uncorrelated configurations from the equilibrium canonical ensemble at a given temperature. From each configuration we do dynamic simulation according to the equations of motion for the XY model, which are given as follows\cite{KL99,HH77}:
\begin{eqnarray}
\frac{\rm d}{{\rm d}t}\vec{S_i}=\frac{\partial {\cal H}}{\partial \vec{S_i}}\times \vec{S_i}.
\end{eqnarray}
Starting from a particular initial spin configuration, we perform numerical integration of these equations of motion using a recently developed decomposition method\cite{FHL97}, which guarantees exact energy conservation and conservation of spin length $|\vec{S_k}|=1$ and conserves $M_z$ within its numerical truncation errors. 
The integration is carried out to a maximum time $t_{max}$ (typically of the order of $t_{max}=200$) to make sure that we determine the real-time history of every configuration within a sufficiently long interval of time ($0\leq t\leq t_{max}$). 
The time step is $\Delta t=0.1$, which guarantees sufficient accuracy with respect to the conservation of $M_z$.
Fourth-order integrators are used, which are more accurate than a second-order method for same time step.
Finally, we compute the average of a time-dependent observable (such as the current-current correlation function) over all results relative to all the independent initial equilibrium configurations generated via the hybrid Monte Carlo procedure.
All error bars for the thermal resistivity correspond to one standard deviation.

\section{\label{sect:results}Results}
In this section, we calculate the thermal resistivity, examine its scaling behavior with respect to $H$, and compare the scaling function with the experimental results.
For a film geometry, $L\times L\times H$, it is better to keep $H$ finite and let $L\rightarrow \infty$ such that the system can not feel the effect of $L$ for any temperature. In other words, we wish the correlation length to be larger than $H$, but always less than $L$ for any temperature. We can realize this by performing extrapolation to the infinite limit from the results on finite $L$. However, we do not have to do this, because we find that by applying periodic boundary conditions along the directions of $L$, the finite-size effects due to $L$ are already not significant as long as $L\geq 5H$ (see Fig. 1) . Fig. 1 shows the thermal resisitivity $R(t,H)=1/ \lambda(t,H)$ for film lattices as a function of $L/H$ at the critical temperature, where the correlation length is infinite. From Fig. 1, we can see that within the error bars, the results of thermal resistivity are same for $L/H \geq 5$.

\begin{figure}
\epsfig{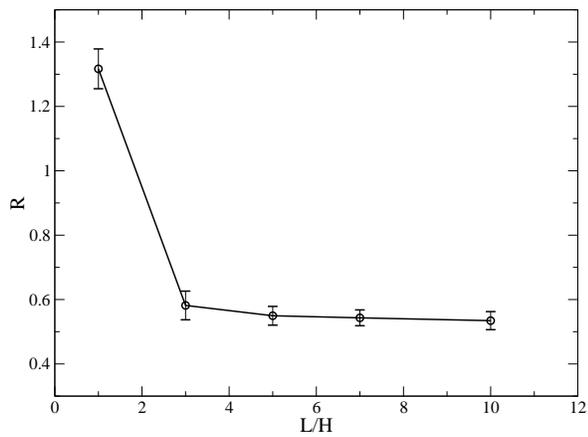}
\caption{
	The thermal resistivity with a fixed $H=6$ at $T_{\lambda}$ as a function of $L/H$.
         }
\label{Fig:conv}
\end{figure}

\begin{figure}
\epsfig{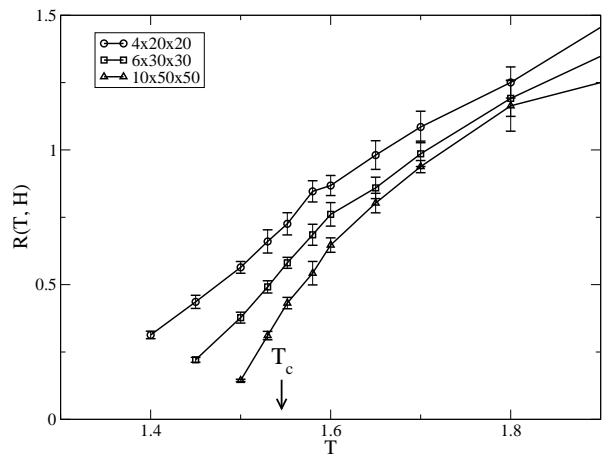}
\caption{
         Thermal resistivity $R(T,H)$ versus temperatures for film geometry with sizes that correspond to $H$ =4, 6, 10 and $L=5H$. 
         The bulk critical temperature $T_{\lambda}=1.5518$ is also shown.
         }
\label{Fig:cresult}
\end{figure}

Fig. \ref{Fig:cresult}  shows some of our results of the thermal resistivity $R(T,H)$ as a function of temperature $T$ for various lattice sizes $H$. The bulk transition temperature $T_{\lambda}=1.5518(2)$ is obtained from Monte Carlo simulation\cite{NM99}. At high temperature, there is no significant scaling behavior and the thermal resistivity for different $H$ has no obvious difference, because at high temperatures, $\xi(t) < H$. Near $T_{\lambda}$, the data show strong effects of confinement. The smallest size shows the greatest rounding of transition and the thermal resistivity is biggest for smallest size at the critical temperature.

Now we would like to check the finite-size scaling hypothesis for the thermal resistivity and compare our results with the existing experimental results\cite{AL05}. The dependence upon $t$ of the bulk thermal resistivity can be described by a power law
\begin{eqnarray}
R(t)=R_0t^{\chi},
\end{eqnarray}
where $\chi$ is the dynamic critical exponent of the thermal resistivity. The dependence upon $t$ of the correlation length can be described by a power law
\begin{eqnarray}
\xi (t) = \xi_0t^{-\nu},
\end{eqnarray}
where $\nu$ is the critical exponent of the correlation length. Using Eq.(1), the finite-size scaling expression for the thermal resistivity $R(t,H)$ is given by
\begin{eqnarray}
R(t,H)H^{\chi / \nu} = g(tH^{1/ \nu})
\end{eqnarray}
where $g(x)$ is a universal function\cite{GA92}.

Fig. 3 shows a scaling plot of the thermal resistivity scaling function $g(x)=R(t,H)H^{\chi/ \nu}$ versus the scaled reduced temperature parameter $x=tH^{1/ \nu}$, where the reduced temperature is taken relative to the bulk transition temperature $T_{\lambda}$. We use the value $\nu =0.6705$ as determined by Goldner and Ahlers\cite{GA92} and $\chi =0.44$ determined by Ahlers\cite{GA99}. From Fig. 3, we can find that near $T_\lambda$ our simulation data collapse onto one single curve for a wide range of values of $H$ and $t$, thus supporting the validity of finite-size scaling theory. When $T$ is large, where $\xi(t) < H$, we can not obtain the unviersal function Eq.(8) using Eq.(1), therefore, the simulation data might not collapse onto one single curve. 
\begin{figure}
\epsfig{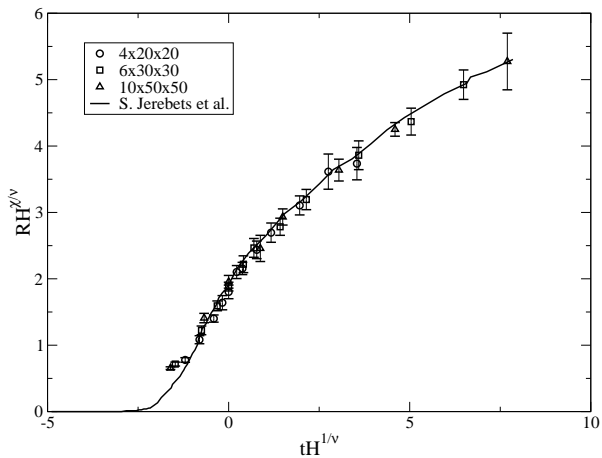}
\caption{
         The universal function $g(x)$ obtained for film geometry. The solid line represents the available experimental results for film geometry. 
In the experimental results the resistivity scale and the temperature scale are used as free parameters.
         }
\label{Fig:match}
\end{figure}

In Fig. 3 we also compare our universal function $g(x)$ with the experimental data obtained by Jerebets et al.\cite{AL05} represented by a solid line. In order to do this, we used two multiplicative constants as free fitting parameters, one multiplying the scale of $x$-axis and another the scale of $y$-axis. The agreement between simulation and experiment is quite good.

The geometry is one of  important factors which, in principle, can determine the scaling function.
 In Fig. 4, we compare the scaling functions for films and pores. 
From Fig. 4 we can find that near $T_\lambda$ the scaling function for films and pores are different alghough their shapes are similar. 
Compared with scaling function for films, the scaling function for pores shifts to the left, further away from the bulk behavior. 
This is expected; since in comparison with the film geometry, the pore geometry restricts the dimensionality of the system more, thereby further limits critical fluctuations. As a consequence the thermal resistivity for pore geometry is higher at $T_{\lambda}$, and decreases to 0 at lower temperature. 

\begin{figure}
\epsfig{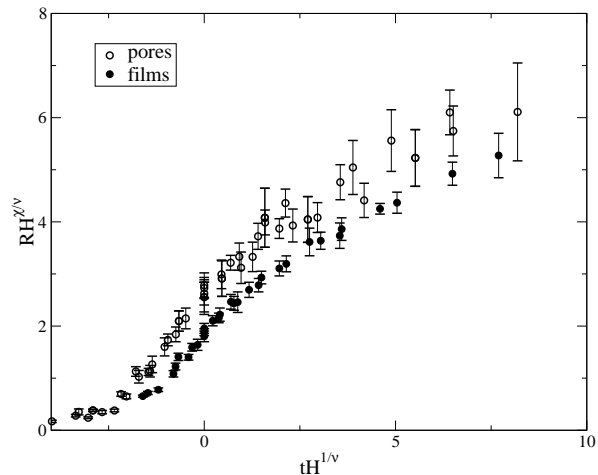}
\caption{
        Comparison of the finite-size scaling functions $g(x)$ for films and pores.
         }
\label{Fig:comp}
\end{figure}

\section{summary\label{sect:summary}}
We have calculated the thermal resistivity $R(t,H)$ of liquid $^4$He in a film geometry (on a $L\times L\times H$ lattice) applying periodic boundary conditions along $L$ directions and open boundary conditions in the $H$ direction. We were able to demonstrate the validity of finite-size scaling theory and we obtained the thermal resistivity scaling function $g(x)$ using known values for the critical exponents and no adjustable parameters. We find good agreement for scaling functions between simulational and experimental results using the temperature scale and the thermal resistivity scale as free parameters. We also compared our calculated scaling function for film geometry with the result for pore geometry and found a systematic shift.

\begin{acknowledgments}
We thank G. Ahlers for helpful comments and suggestions. This work was supported by the NASA under grant No. NNC04GB24G.
\end{acknowledgments}

\bibliography{p1}

\end{document}